\documentclass[twocolumn,superscriptaddress,prl]{revtex4}
\usepackage{graphicx}
\usepackage{epsfig}
\usepackage{color}
\usepackage{amsmath,amssymb,subfigure,psfrag}
\def\U#1{{\rm #1}} 
\def\u#1{_{\rm #1}}

\newcommand{\bra}[1]{\langle #1 |}
\newcommand{\ket}[1]{| #1 \rangle}

\newcommand{\dagg}[1]{#1 ^\dagger}

\def\H{{\rm H}}
\def\V{{\rm V}}
\def\tr{\U{tr}}

\def\Dv{\U{D}\u{v}}
\def\Dt{\U{D}\u{t}}

\begin{document}
\title{
High-fidelity conversion of photonic quantum information 
to telecommunication wavelength 
with superconducting single-photon detectors
}

\author{Rikizo~Ikuta}
\affiliation{Graduate School of Engineering Science, Osaka University,
Toyonaka, Osaka 560-8531, Japan}
\author{Hiroshi~Kato}
\affiliation{Graduate School of Engineering Science, Osaka University,
Toyonaka, Osaka 560-8531, Japan}
\author{Yoshiaki~Kusaka}
\affiliation{Graduate School of Engineering Science, Osaka University,
Toyonaka, Osaka 560-8531, Japan}
\author{Shigehito~Miki}
\affiliation{
Advanced ICT Research Institute, 
National Institute of Information and Communications Technology (NICT), 
588-2 Iwaoka, Kobe 651-2492, Japan}
\author{Taro~Yamashita}
\affiliation{
Advanced ICT Research Institute, 
National Institute of Information and Communications Technology (NICT), 
588-2 Iwaoka, Kobe 651-2492, Japan}
\author{Hirotaka~Terai}
\affiliation{
Advanced ICT Research Institute, 
National Institute of Information and Communications Technology (NICT), 
588-2 Iwaoka, Kobe 651-2492, Japan}
\author{Mikio~Fujiwara}
\affiliation{
Advanced ICT Research Institute, 
National Institute of Information and Communications Technology (NICT), 
4-2-1 Nukuikawa, Koganei, Tokyo 184-8795, Japan}
\author{Takashi~Yamamoto}
\affiliation{Graduate School of Engineering Science, Osaka University,
Toyonaka, Osaka 560-8531, Japan}
\author{Masato~Koashi}
\affiliation{Photon Science Center, The University of Tokyo, 
Bunkyo-ku, 113-8656, Japan}
\author{Masahide~Sasaki}
\affiliation{
Advanced ICT Research Institute, 
National Institute of Information and Communications Technology (NICT), 
4-2-1 Nukuikawa, Koganei, Tokyo 184-8795, Japan}
\author{Zhen~Wang}
\affiliation{
Advanced ICT Research Institute, 
National Institute of Information and Communications Technology (NICT), 
588-2 Iwaoka, Kobe 651-2492, Japan}
\author{Nobuyuki~Imoto}
\affiliation{Graduate School of Engineering Science, Osaka University,
Toyonaka, Osaka 560-8531, Japan}

\begin{abstract}
We experimentally demonstrate a high-fidelity 
visible-to-telecommunication wavelength conversion 
of a photon 
by using a solid-state-based difference frequency generation. 
In the experiment, 
one half of a pico-second visible entangled photon pair at 780 nm 
is converted to a 1522-nm photon, 
resulting in the entangled photon pair between 780 nm and 1522 nm. 
Using superconducting single-photon detectors 
with low dark count rates and small timing jitters, 
we selectively observed well-defined temporal modes 
containing the two photons. 
We achieved a fidelity of $0.93 \pm 0.04$ 
after the wavelength conversion, 
indicating that our solid-state-based scheme can be used 
for faithful frequency down-conversion 
of visible photons emitted from quantum memories 
composed of various media. 
\end{abstract}

\maketitle

Wavelength conversion of photons 
in a quantum regime~\cite{conv} 
has been actively studied~\cite{tanzilli, eraser, 
matthew, McGuinness, four2, ikuta, up, arXiv1, arXiv2} 
as a quantum interface 
for application of quantum information processing 
and communications. 
Especially, such a conversion 
aiming at 
near-infrared photons in telecommunication bands 
are essential for transmitting quantum information 
over long-distance optical fiber networks 
with quantum repeaters~\cite{communication, Rep1, Rep2}. 
In the quantum repeaters, 
the photon sent to a relay point through an optical fiber 
needs to be entangled with a quantum memory. 
At present, 
many of quantum memories 
and processors based on alkaline atoms, 
trapped ions and solid states have successfully 
created entanglement with photons 
at around visible wavelengths~\cite{Rb1, Rb2, Cs1, Yb1, Nv1, rempe, pro1}. 
Thus, a quantum interface for the wavelength conversion 
from visible to telecommunication bands 
with a high fidelity has attracted much interest 
for its applications. 
So far such a quantum interface has been demonstrated 
by using four wave mixing with a cold atomic cloud~\cite{four2} 
or difference frequency generation~(DFG) 
from a nonlinear optical crystal~\cite{ikuta}. 
Among them, nonlinear optical crystals 
with waveguide structure have practically desirable features. 
They can operate near room temperature 
and do not require laser cooling configuration, 
enabling a compact setup and integration 
into a photonic quantum circuit 
on a chip using waveguide structures~\cite{chip}. 
In addition, 
they have a wider bandwidth, 
compatible with 
wide-band quantum memories~\cite{waveguidememory, warm}, 
resulting in high-clock-rate quantum information processing. 
Such kind of solid-state-based optical quantum interface lead 
to development of a mature quantum information technology. 
However, in several demonstrations 
of the solid-state-based 
wavelength conversion~\cite{ikuta,DFG1, DFG2, matthew,noise1}, 
they suffered from degradation of an observed fidelity of 
a reconstructed quantum state after the wavelength conversion 
due to background noises caused 
by Raman scattering of a strong cw pump light 
and relatively high dark count rate 
of an InGaAs/InP avalanche photodiode~(APD) 
for photon detection at the telecommunication band. 
Therefore the observed fidelity of the state 
after the wavelength conversion 
is degraded. 

In this Letter, 
we demonstrate almost noiseless wavelength conversion 
by suppressing the effects of 
both the optical noise from the Raman scattering 
and the dark count 
via newly developed superconducting single-photon 
detectors~(SSPDs) 
for visible and telecommunication wavelengths 
of the photons~\cite{NICT1, NICT2}. 
The SSPDs have lower dark count rates 
and smaller timing jitters than those of typical APDs. 
Especially, 
the latter property enables us 
to selectively observe 
well-defined temporal modes containing the two photons. 
Because duration of signal photons 
in our experiment is of pico-second order 
whereas the optical noise 
through the wavelength conversion are continuously generated, 
the use of the SSPDs will lead 
to reduction of irrelevant photon detections. 
The observed fidelity of the two-photon state 
after the wavelength conversion 
to a maximally entangled state is $0.93 \pm 0.04$, 
which is very close to the initial fidelity 
of $0.97\pm 0.01$. 
We also clearly observe 
the violation of the Clauser-Horne-Shimony-Holt-type Bell's 
inequality with $S = 2.62 \pm 0.09$. 

Theoretical treatment of wavelength conversion 
of a single mode of a pulsed light is as follows~\cite{conv, ikuta}. 
When a pump light at angular frequency $\omega\u{p}$ 
is sufficiently strong, 
the Hamiltonian of the wavelength conversion 
using a second-order nonlinear optical interaction 
is described by 
${\hat{H}}=i\hbar \sqrt{\eta P}
(e^{-i\varphi} \dagg{\hat{a}}\u{c}\hat{a}\u{s}
- e^{i\varphi} \dagg{\hat{a}}\u{s}\hat{a}\u{c})$. 
Here 
$\hat{a}\u{s}$ and $\hat{a}\u{c}$ are annihilation operators of 
a signal mode at angular frequency $\omega\u{s}$ 
and a converted mode 
at angular frequency $\omega\u{c}=\omega\u{s} - \omega\u{p}$, 
respectively. $P$ and $\varphi$ 
are a power and a phase of the classical pump light, respectively. 
$\eta$ is a constant 
and $\sqrt{\eta P}e^{i\varphi}$ represents 
an effective coupling constant. 
Using this Hamiltonian, 
an annihilation operator $\hat{a}\u{c,out}$ 
of the converted mode coming from the nonlinear optical crystal 
is described by using the time evolution of 
$\hat{a}\u{s}$ in the Heisenberg picture as 
\begin{eqnarray}
\hat{a}\u{c,out}=\hat{a}\u{c}(\tau)
= 
e^{-i\varphi}
\sin(\sqrt{\eta P} \tau)\ \hat{a}\u{s} 
+ \cos(\sqrt{\eta P} \tau)\ \hat{a}\u{c}, 
\label{as}
\end{eqnarray}
where $\tau$ is the traveling time of the pulses 
through the crystal. 

\begin{figure}
 \begin{center}
 \scalebox{0.5}{\includegraphics{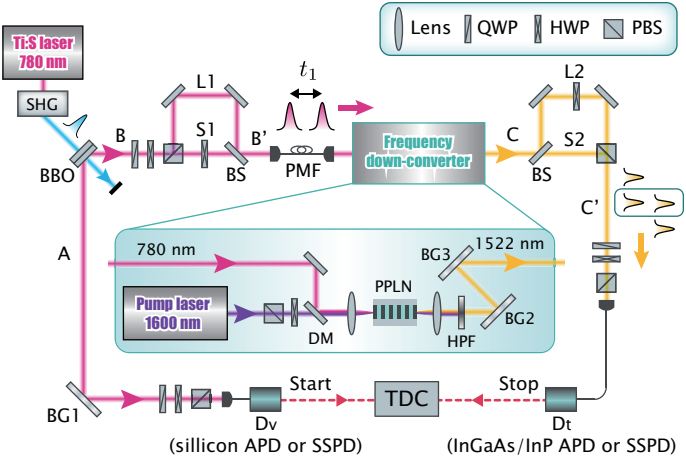}}
  \caption{
  The experimental setup for frequency down-conversion 
  of one halves of the visible entangled photon pairs. 
  \label{fig:setup}}
 \end{center}
\end{figure}
The experimental setup 
for the DFG-based frequency down-conversion 
of one half of 
a polarization-entangled photon pair at 780 nm 
to the wavelength of 1522 nm is shown in Fig.~\ref{fig:setup}. 
We use a mode-locked Ti:sapphire laser~(wavelength: 780 nm; 
pulse width: 1.2 ps; repetition rate: 82 MHz) as a light source. 
It is frequency doubled by a second harmonic generator~(SHG), 
and then the UV pulse with a power of 250 mW 
pumps a pair of type-I 
phase-matched $1.5$-mm-thick $\beta$-barium borate~(BBO) crystals 
to prepare the 
polarization-entangled photon pair A and B 
which is described as 
$\ket{\phi^+}\u{AB}\equiv 
(\ket{\H\H}\u{AB}+\ket{\V\V}\u{AB})/\sqrt{2}$ 
through spontaneous parametric down-conversion~\cite{DFS}. 
Here $\ket{\H}$ and $\ket{\V}$ 
represent horizontal~(H) and vertical~(V) 
polarization states of a photon. 
The spectrum of photon A is narrowed 
by a Bragg grating~(BG1) with a bandwidth of 0.2 nm, 
and then the photon is detected by detector $\Dv$ 
connected to a single-mode fiber. 
We switch the silicon APD and the SSPD for $\Dv$. 
Photon B is split into a short path~(S1) 
and a long path~(L1) according to a polarization of the photon. 
A half-wave plate~(HWP) in S1 flips the polarization from H to V. 
As a result, a polarization qubit~$\{ \ket{\H}, \ket{\V}\}$ 
in mode B is transformed to a time-bin 
qubit~$\{ \ket{\U{S1}}, \ket{\U{L1}}\}$ in mode $\U{B'}$, 
leading to a two-photon state in modes A and $\U{B'}$ of 
$\ket{\psi}\u{AB'} \equiv 
(\ket{\H}\u{A}\ket{\U{S1}}\u{B'}
+\ket{\V}\u{A}\ket{\U{L1}}\u{B'})/\sqrt{2}$. 
Here $\ket{\U{S1}}$ and $\ket{\U{L1}}$ represent 
states of V-polarized photons passing through 
S1 and L1, respectively. 
We set the time difference between S1 and L1 to about 700 ps. 
The photon in mode $\U{B'}$ goes through 
a polarization-maintaining fiber~(PMF) 
and then enters the frequency down-converter 
whose details are shown in the inset of Fig.~\ref{fig:setup}. 

For the DFG of the signal photon at 780 nm, 
a cw pump laser at 1600 nm is used. 
The linewidth of the pump light is 150 kHz, 
and its coherence time is much longer than 
the time difference of the photons passing through S1 and L1. 
The pump light is combined with the signal photon 
at a dichroic mirror~(DM) 
after its polarization is set to V 
by a polarization beamsplitter~(PBS) and a HWP. 
Then they are focused on the type-0 quasi-phase 
matched~($\V \rightarrow \V+ \V$) PPLN waveguide~\cite{nishikawa} 
whose temperature is controlled to be about 50$^\circ$C. 
The length of the PPLN crystal is 20 mm and 
the acceptable bandwidth is calculated to be about 0.3 nm. 
After passing through the PPLN waveguide, 
the strong pump light is diminished 
by a high-pass filter~(HPF), 
and the light converted to the wavelength of 1522 nm 
is extracted by BG2 and BG3 whose bandwidths are 1 nm. 

The photon in mode C from the frequency down-converter is 
split into a short path~(S2) and a long path~(L2) by a BS. 
The polarization of the photon passing through L2 
is flipped from V to H by a HWP. 
Time difference between S2 and L2 
is adjusted to be the same as that between S1 and L1. 
The components of the photon from S2 and L2 
are recombined by a PBS, and the photon in mode $\U{C'}$ 
is detected by $\Dt$ 
after passing through a single-mode fiber. 
We switch the InGaAs/InP APD and the SSPD for $\Dt$. 

Each SSPD consists of 
an 100-nm-thick Ag mirror, a $\lambda/4$ SiO cavity 
and 4-mm-thick niobium nitride meander nanowire 
on a 0.4-mm-thick MgO substrate from the top. 
The nanowire is 80-nm-wide, 
and it covers an area of 15 $\mu$m $\times$ 15 $\mu$m. 
The respective optical cavity structures of the SSPDs for $\Dv$ and $\Dt$ 
are designed for visible and the telecommunication wavelengths 
to achieve higher detection efficiencies. 
The detection efficiencies are $32$ \% and 
$12.5$ \% for 780 nm and 1522 nm wavelengths, respectively. 
Each of the SSPD chip is shielded by a copper block 
which has a holder of a single-mode optical fiber 
followed by a small-gradient index lense for efficient coupling. 
The blocks are installed 
in the Gifford-McMahon cryocooler system 
whose cooling temperature is $2.28\pm 0.02$ K. 

To measure the coincidence events 
of the detections at $\Dv$ and $\Dt$, 
signals from $\Dv$ and $\Dt$ 
are input to a time-to-digital converter~(TDC) 
as a start and stop signals of a clock, respectively. 
By post-selecting the events 
where the photon in mode $\U{C'}$ has passed through S1-L2 or L1-S2, 
we obtain the polarization-entangled state 
$\ket{\phi^+}\u{AC'}$ in modes A and $\U{C'}$. 
In our experiment, we accept such events 
in 200-ps time window. 

\begin{figure}
 \begin{center}
 \scalebox{0.9}{\includegraphics{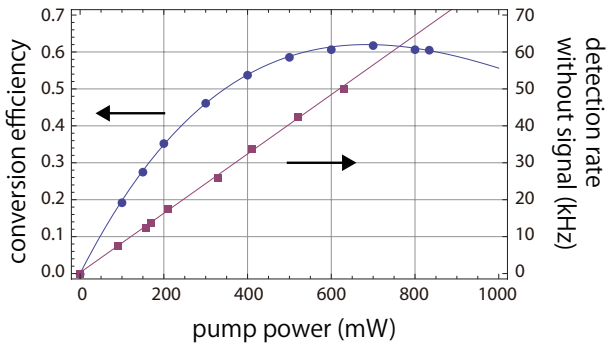}}
  \caption{
  The dependencies of the conversion efficiency and 
  the rate of the background noises on the pump power. 
  The former has been measured 
  by using a pico-second coherent light in Ref.~\cite{ikuta}. 
  The latter was measured by using the SSPD 
  for $\Dt$ without the UV pulse. 
  The first vertical axis is the conversion efficiency. 
  A maximum conversion efficiency 
  is achieved at a pump power of 700 mW. 
  The second vertical axis is detection rate of 
  the background noises. 
  We fitted the experimental data by a function of $b P + d$. 
  $b$ and $d$ are estimated as 80 Hz/mW and 266 Hz, respectively. 
  \label{fig:raman}}
 \end{center}
\end{figure}
For a faithful wavelength conversion, 
the rate of background noises 
must be sufficiently small 
compared to that of the converted photons. 
The background noises are mainly 
caused by the Raman scattering of the pump light 
and dark countings of a photon detector. 
The former is proportional to the pump power and 
is written by $bP$ with a constant $b$. 
The latter is described by a constant $d$. 
Taking into account these noise effects 
and using Eq.~(\ref{as}), 
the signal-to-noise ratio for the observed converted photons 
is represented by 
\begin{eqnarray}
f\u{SNR}(P)\equiv \frac{a \sin^2(\sqrt{\eta P}\tau)}{bP+d}, 
\end{eqnarray}
where $a$ is a constant. 
When $d$ is comparable to $bP$ 
as in the case where 
both detectors $\Dv$ and $\Dt$ are APDs~\cite{ikuta}, 
a maximum of $f\u{SNR}(P)$ is achieved 
near a pump power $P\u{max}$
which gives the maximum conversion efficiency, 
and the maximum of $f\u{SNR}(P)$ 
is close to 
$f\u{SNR}(P\u{max})=a/(bP\u{max}+d)$. 
On the other hand, 
when $d$ is sufficiently small 
as in the case of using the SSPDs, 
we may improve $f\u{SNR}(P)$ significantly 
by decreasing the pump power. 
Using $d/b\approx 3.3$ mW in Fig.~\ref{fig:raman} 
and $\eta \tau^2 \approx 3.6\ \U{W}^{-1}$\cite{ikuta}, 
we see that $f\u{SNR}(P)$ reaches its maximum 
when the pump power is decreased 
from $P\u{max}\approx 700$~mW to $50$ mW. 
In the following experiments, 
we chose the pump power to be 160 mW, 
for which the conversion efficiency 
is not severely degraded~(about half the maximum value) 
and the signal-to-noise ratio is expected to be above 10. 

\begin{figure}
 \begin{center}
 \scalebox{0.7}{\includegraphics{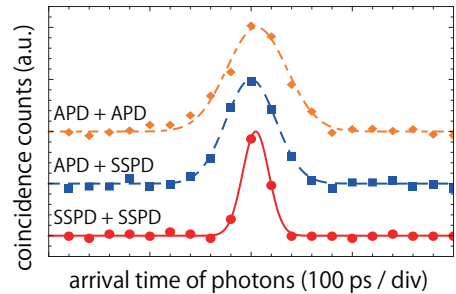}}
  \caption{
  The coincidence counts of the photon pairs 
  recorded by the TDC with the timing resolution of 100 ps 
  when we used the pair of 
  the silicon APD for $\Dv$ and InGaAs/InP APD for $\Dt$~(rhombus), 
  the silicon APD for $\Dv$ and the SSPD for $\Dt$~(square), 
  and the SSPDs for both the detectors~(circle). 
  Each result of the coincidence counts was fitted 
  by the Gaussian function 
  after the subtraction 
  of the counts from the background noises. 
  \label{fig:jitter}}
 \end{center}
\end{figure}
Before the wavelength conversion of the entangled photon pairs, 
we measured the variance 
of jitter in the arrival time of pico-second photons 
to see the high timing resolution of the SSPDs. 
When we used the APDs for both detectors, 
the full width at half maximum~(FWHM) 
of the time distribution of the coincidence counts 
was measured to be $\approx 350$ ps. 
When we replaced the InGaAs/InP APD with the SSPD for $\Dt$, 
the FWHM was measured to be $\approx 290$ ps. 
When we used the SSPDs for both detector, 
the FWHM became $\approx 150$ ps. 
These results which are shown in Fig.~\ref{fig:jitter} 
clearly show that 
the SSPDs gather the coincidence counts 
in the smaller time bins. 
The use of the SSPDs instead of the APDs 
is expected to improve 
the ratio of the signal photons 
to the optical noises by a factor of about $1.8$ 
in our case of 200-ps time window. 

\begin{figure}
 \begin{center}
 \scalebox{0.4}{\includegraphics{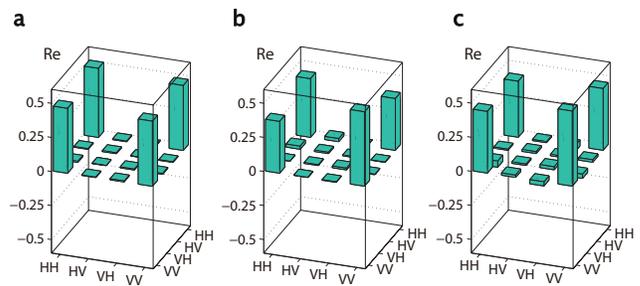}}
  \caption{
  The real parts of the reconstructed density matrices. 
  (a) The initial entangled photon pair $\rho\u{AB}$ 
  prepared by the BBO crystal. 
  (b) The two-photon state $\rho^{\U{AS}}\u{AC'}$ 
  after the wavelength conversion 
  when we used the silicon APD for $\Dv$ and 
  the SSPD for $\Dt$. 
  (c) The two-photon state $\rho^{\U{SS}}\u{AC'}$ 
  after the wavelength conversion
  when we used the SSPDs for both detector 
  $\Dv$ and $\Dt$. 
  \label{fig:matrix}}
 \end{center}
\end{figure}
\begin{table}
\begin{center}
\begin{tabular*}{8.5cm}
{@{\extracolsep{\fill}}c|cccc}
& $F$ & EOF & purity & $S$ \\ 
\hline
$\rho\u{AB}$ & 
$0.97\pm 0.01$ & $0.97\pm 0.03$ & $0.97\pm 0.02$ & $2.73\pm 0.02$\\ 
$\rho^{\U{AS}}\u{AC'}$ & 
$0.87\pm 0.06$ & $0.68\pm 0.15$ & $0.79\pm 0.08$ & $2.35\pm 0.10$\\ 
$\rho^{\U{SS}}\u{AC'}$ & 
$0.93\pm 0.04$ & $0.88\pm 0.10$ & $0.93\pm 0.07$ & $2.62\pm 0.09$\\ 
$\rho'\u{AC'}$ \cite{ikuta}& 
$0.75\pm0.06$ & $0.36\pm 0.13$ & - & - 
\end{tabular*}
 \caption{
 The observed fidelities, the EOFs the purities, and 
 the $S$ parameters of the reconstructed operators 
 before and after the wavelength conversion. 
 $\rho'\u{AC'}$ is a reconstructed operator 
 of photons in modes A and $\U{C}'$ 
 by using the silicon APD and the InGaAs/InP APD 
 for $\Dv$ and $\Dt$, respectively~\cite{ikuta}. 
 The attached errors are the standard deviations~($1$-$\sigma$) 
 with the assumption of the Poisson statistics of the counts. 
 \label{tbl:results}}
\end{center}
\end{table}
In the experiment of the frequency down-conversion 
of the visible entangled photon pairs, 
we first performed quantum state tomography of 
the initial photon pairs in modes A and B at 780 nm 
by rotating a quarter-wave plate~(QWP) and a HWP 
followed by a PBS in each mode~\cite{MQ}. 
We used a silicon APD for $\Dv$. 
The PMF before the frequency down-converter 
was connected to a silicon APD for detection of photons in mode B. 
We observed the two-photon state in modes A and B 
with a detection rate of 444 Hz. 
Using the iterative maximum likelihood method~\cite{MLE1}, 
the density operator $\rho\u{AB}$ was reconstructed 
as shown in Fig.~\ref{fig:matrix}~(a). 
From the reconstructed $\rho\u{AB}$, 
we calculated the fidelity defined by $\bra{\phi^+}\rho\u{AB}\ket{\phi^+}$, 
the entanglement of formation~(EOF)~\cite{wooters}, 
and the purity defined by $\tr(\rho\u{AB}^2)$ 
as $0.97\pm 0.01$, $0.97\pm 0.03$ and $0.97\pm 0.02$, respectively. 
These results show that 
the 780-nm photon pair 
before the wavelength conversion was highly entangled. 
Next, we performed quantum state tomography 
of the photon in mode A and the converted photon in mode $\U{C'}$ 
when we used the silicon APD for $\Dv$ and the SSPD for $\Dt$. 
An observed detection rate of the two photons was 
$0.324$ Hz including background noises at a rate of $0.039$ Hz. 
The reconstructed density operator 
$\rho^{\U{AS}}\u{AC'}$ is shown in Fig.~\ref{fig:matrix}~(b). 
The fidelity, EOF and the purity 
are shown in Table~\ref{tbl:results}. 
We then switched the silicon APD to the SSPD for $\Dv$, 
and performed quantum state tomography. 
In this case, we detected the two photons after the conversion 
at a rate of $0.280$ Hz 
including noise photons at a rate of $0.015$ Hz. 
The reconstructed density operator 
$\rho^{\U{SS}}\u{AC'}$ is shown in Fig.~\ref{fig:matrix}~(c). 
The observed fidelity, EOF and the purity are shown in Table~\ref{tbl:results}. 
We also performed the tests of the 
Clauser-Horne-Shimony-Holt-type Bell's inequality~\cite{CHSH} 
by observing the $S$ parameter. 
It is known that any local hidden variable theory leads to $S\leq 2$. 
The experimental 
results of the $S$ parameters are shown in Table~\ref{tbl:results}. 
The Bell's inequality was clearly violated with over $6$-$\sigma$ deviation. 
From these results, we see 
that the high-fidelity frequency 
down-conversion was achieved, 
which could be more clearly observed with SSPDs. 

In conclusion, 
we have demonstrated the faithful solid-state-based 
frequency down-conversion of visible photons at 780 nm 
to a telecommunication wavelength of 1522 nm. 
Thanks to the lower dark count rates 
and the smaller timing jitters of the SSPDs, 
the observed fidelity of the two-photon state 
after the wavelength conversion was $0.93\pm 0.04$, 
which was very close to the value of the fidelity 
obtained before the conversion. 
In this experiment, 
while the suppression of the effect of the Raman scattering 
for the faithful wavelength conversion 
was achieved at the cost of the conversion efficiency, 
it will be attained without decreasing the conversion efficiency 
by a faster time-resolution  measurement. 
Our demonstration shows the possibility of a 
noiseless and wide-band solid-state-based frequency down-conversion. 
We believe that such a quantum interface 
is vital for building quantum networks 
based on repeaters 
and for performing various quantum communication protocols. 

This work was supported by the Funding Program for 
World-Leading Innovative R \& D on Science and 
Technology (FIRST), MEXT Grant-in-Aid for Scientific 
Research on Innovative Areas 20104003 and 21102008, 
the MEXT Global COE Program, and 
MEXT Grant-in-Aid for Young scientists(A) 23684035.

\end{document}